\documentclass[iop]{emulateapj}

\usepackage{txfonts}
\usepackage{graphicx}
\usepackage{natbib}

\def\etal{{\sl et al.}}
\usepackage{natbib}
\shorttitle{Ellerman Bomb association with Chromospheric Jet Foot-points}
\shortauthors{Reid et al.}

\begin{document}

\title{ELLERMAN  BOMBS WITH JETS: CAUSE AND EFFECT}
\vskip1.0truecm
\author{
A. Reid$^{1,2}$, M. Mathioudakis$^{1}$, E. Scullion$^{3}$, J. G. Doyle${^2}$, S. Shelyag${^4}$, P. Gallagher${^3}$}
\affil{
1. Astrophysics Research Centre, School of Mathematics and Physics, Queen's University Belfast, BT7~1NN, Northern Ireland, UK; e-mail: areid29@qub.ac.uk\\
2. Armagh Observatory, College Hill, Armagh, BT61 9DG, UK\\
3. School of Physics, Trinity College Dublin, Dublin 2, Ireland.\\
4. Monash Centre for Astrophysics, School of Mathematical Sciences, Monash University, Clayton, Victoria, 3800, Australia\\}
%

\begin{abstract}
Ellerman Bombs (EBs) are thought to arise as a result of photospheric magnetic reconnection. We use data from the Swedish 1-m Solar Telescope (SST), to study EB events on the solar disk and at the limb. Both datasets show that EBs are connected to the foot-points of forming chromospheric jets. The limb observations show that a bright structure in the H$\alpha$ blue wing connects to the EB initially fuelling it, leading to the ejection of material upwards. The material moves along a loop structure where a newly formed jet is subsequently observed in the red wing of H$\alpha$. In the disk dataset, an EB initiates a jet which propagates away from the apparent reconnection site within the EB flame. The EB then splits into two, with associated brightenings in the inter-granular lanes (IGLs). Micro-jets are then observed, extending to  500~km with a lifetime of a few minutes. Observed velocities of the micro-jets are approximately 5-10 km s$^{-1}$, while their chromospheric counterparts range from 50-80 km s$^{-1}$. MURaM simulations of quiet Sun
reconnection show that micro-jets with similar properties to that of the observations follow the line of reconnection in the photosphere, with associated H$\alpha$ brightening at the location of increased temperature. 

\end{abstract}


\section{INTRODUCTION}

Ellerman Bombs (EBs) are usually identified as prominent small-scale brightenings in the wings of the H$\alpha$ line \citep{ell}. They have mean lifetimes of $\sim$10-15 minutes and are mainly observed in the vicinity of active regions \citep{zac, geo}, or regions of emerging magnetic flux \citep{iso, wat}. The absorption core of H$\alpha$ remains unchanged, relative to a background local line profile, at the EB location. The wing enhancements can often be asymmetric as a result of Doppler shifts due to overlying chromospheric flows \citep{kitai}. It is thought that EBs produce no observable effect in the transition region and corona \citep{viss}. However, \cite{schmieder} found increased EB activity beneath brighter ``moss'' areas, hinting at a possible contribution to the heating of the transition region, with no substantial evidence for any coronal effects.
 
The correct classification of EBs is paramount for identifying the driving mechanism(s) of these events. A recent review of EBs \citep{rutt} emphasises that not all H$\alpha$ wing brightenings should be classified as EBs. Some of the previous research is now being questioned on this premise, creating a potential minefield of inconsistencies.
 
Recent studies show a connection between EBs and opposite polarity photospheric magnetic fields indicating that  photospheric reconnection is the driving mechanism for EBs \citep{Nelson, Nelson1, viss, geo, wat, matsu, hashi}. It has been shown numerically that photospheric magnetic reconnection would be most efficient at the temperature minimum at a height of 600~km above the lower photospheric boundary \citep{lit2}. EBs have also been observed to have some structuring, with heights ranging from 600~km \citep{wat1} and up to 1300~km \citep{zac}, generally seen together with blue-shifts, or bi-directional Doppler shifts \citep{wat1, matsu2}. The apparent structuring of EBs with height is supported by the detection of these events in SDO 1600~\AA\, and 1700~\AA\ continuum datasets \citep{viss}.

\begin{figure*}[!t]
\begin{center}
\plotone{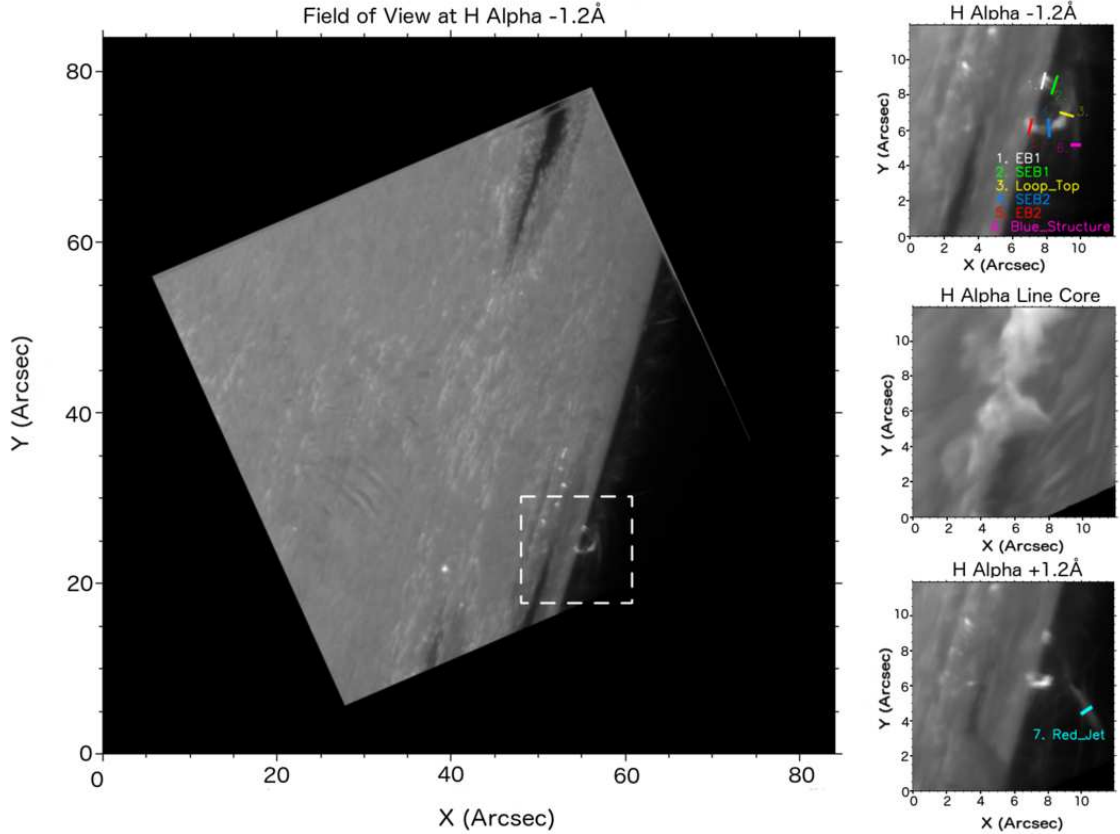}
\end{center}
\centering
\caption{Left Panel: The full FOV with a white box indicating the region of interest. Right Panels: Co-temporal and co-spatial imaging of the loop structure and EBs at the limb, in H$\alpha$ line core (center) and $\pm 1.2$~\AA\ from  line core. A movie available online shows the co-temporal evolution of the system at -1.2~\AA\ and +1.2~\AA\ from the line core. The cut-off at the bottom right of the images is due to the edge of the CRISP FOV. These frames are taken at time T = 1800s from Movie1.}
\label{Figure1}
\end{figure*}

It has been proposed that EBs may be triggered by 3 main mechanisms with associated magnetic topologies. The first is photospheric reconnection triggered by the interaction of new emerging flux with an existing and opposite polarity magnetic topology \citep{wat, hashi}. In the second process, EBs can be triggered in a unipolar field, with shearing field lines that reconnect \citep{geo, wat}. The third scenario involves the reconnection of a resistive, undulatory ``sea serpent" flux emergence \citep{geo, par}, which has also been studied numerically \citep{ach,lit2}.

 \cite{ach} used three-dimensional numerical modelling to show a co-spatial increase in temperature of the lower solar atmosphere with a magnetic topology similar to the ``sea serpent" case. A localised increase in temperature occurs between opposite polarity fields. This temperature increase would in turn increase the flux in the H$\alpha$ line wings, the primary EB signature. Semi-empirical models of the solar atmosphere also show that a temperature increase in the upper photosphere/lower chromosphere leads to intensity enhancements in the wings of the H$\alpha$ and Ca II 8542~\AA\ lines \citep{fang}. These findings have been reinforced by the recent observations and NLTE modelling of \cite{berlicki}. MURaM simulations show that these lower  atmospheric local temperature enhancements are produced at the reconnection sites \citep{Nelson1} and are co-spatial with Fe I 6302\AA\ line core and H$\alpha$ wing intensity increases. More recently, \cite{hong} used a two-cloud model to describe EBs, with the photospheric atmosphere in one cloud, and the overlying chromospheric canopy in the other. This allowed for the fitting of the observed H$\alpha$ and Ca II 8542~\AA\ line profiles,  showing a temperature increase of 400-1000K relative to the quiet sun in the lower photospheric cloud.

A connection between EBs and surges in H$\alpha$ has also been observed in some studies (e.g. \citep{matsu, wat1, yang}). \cite{roy} observed that 56\% of EBs had associated ejections at near disk-center, with 86\% of near limb EBs showing related ejections. These surges generally occur a few minutes after the appearance of the EB, with Doppler velocities ranging from 20 - 100 km s$^{-1}$ \citep{wat1, yang, roy}. These surges exist for roughly 60 - 300 seconds, with opposite Doppler shifts occasionally being noted afterwards \citep{wat1, yang}. 

With the advent of high resolution instruments on ground-based and space-borne facilities, detailed studies of EBs have been made possible. In this paper, we study the link between EBs and chromospheric jets. The datasets under investigation include EBs at disk centre and at the solar limb, with chromospheric jets a common feature for both. Co-spatial and co-temporal high resolution magnetograms allow us to follow the evolution of the on-disk EB in tandem with the magnetic field. Our results are compared with simulations of radiative MHD. 

\section{OBSERVATIONS AND DATA REDUCTION}

The observations were carried out with the CRisp Imaging SpectroPolarimeter (CRISP) at the Swedish 1-m Solar Telescope \citep{sst, sst2} on La Palma.  
 The first target was active region NOAA 11504 consisting of 2 sunspots at the limb (Coordinates: X= 893, Y= -250 ). NOAA 11504 was observed in H$\alpha$ on 2012 June 21 at 07:18-07:48 UT using 31 equally spaced line positions with steps of 86~m\AA, from -1.376~\AA\ to +1.29~\AA\ relative to line centre, with an additional 4 positions in the far blue wing from -1.376~\AA\ to -2.064~\AA.  This dataset has also been studied by \cite{Nelson3},  though not the event analysed in this paper. The SST data were combined with co-aligned data from various passbands of the Atmospheric Imaging Assembly (AIA) \citep{lemen} on the Solar Dynamics Observatory (SDO). 

The second observation was of active region NOAA 11857 and was carried out on 2013 October 07 from 09:38-10:46 UT   (Coordinates: X= -137, Y= -238). The CRISP spectra comprised of 11 positions across the H$\alpha$ line profile equally spaced between -1.2~\AA\ to +1.2~\AA\ from line centre in steps of 258~m\AA. This dataset also contains simultaneous single line position Fe I 6302~\AA\ polarimetry data, -0.04~\AA\ from line core, to obtain the Stokes I, Q, U, and V parameters. This position was chosen as it corresponds to the peak of the Stokes-$V$ signal in the  6302~\AA\ FeI line of small magnetic concentrations in quiet sun \citep{pont2, hewitt}. A single line position   allows the acquisition of magnetograms  without a significant reduction in the cadence of the H$\alpha$ scan. We emphasize that this  technique is sufficient for context magnetograms but does not provide values for the magnetic field strength. Both datasets have an overall cadence of 8 seconds and an image scale of 0.0592$''$ per pixel.

The data were processed using the Multi-Object Multi-Frame Blind Deconvolution (MOMFBD) algorithm \citep{noort}. This includes tessellation of the images into 64x64 pixels$^2$ sub-images for individual restoration, done over each temporal frame and line position within the scans. Wide-band images, in each dataset, act as a stabiliser for the narrow-band alignment. Prefilter Lorentzian corrections are applied to the restored images. 
The final correction involves the long-scale cavity error of the instrument.  Further information on MOMFBD image restoration techniques is available in \cite{noort2} and \cite{noort3}. For the purpose of this paper the definition of an Ellerman Bomb was taken to be a 150\% brightening in the H$\alpha$ wings, relative to a background profile, with no brightening in the core (based on 140-155\% threshold from \citep{viss}), along with evidence of structure flaring \citep{wat1, viss}.\\

\section{RESULTS AND DISCUSSION}
\subsection{Ellerman Bombs at the limb}

\begin{figure}[!h]
\plotone{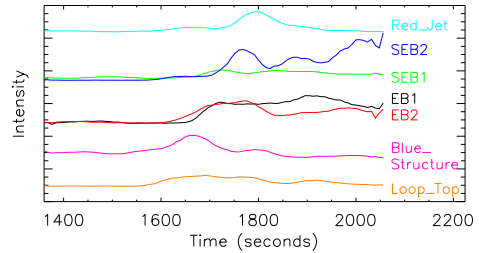}

\caption{Lightcurves generated from the areas highlighted in Fig.~\ref{Figure1}. The scaling of each lightcurve has been changed to reduce overlap where appropriate.}
\label{Figure2}
\end{figure}

 In Fig.~\ref{Figure1} we show a snapshot of the full field-of-view (FOV) from the limb dataset, along with three images across the H$\alpha$ line profile of the region  of interest. A small loop structure is visible (a movie covering the sub-FOV of this dataset is available online, named Movie1, which covers times from T=1440-2064s with T=0s corresponding to the beginning of the observations). A brightening is seen in the H$\alpha$ blue wing parallel to the limb, impacting at the loop apex (labelled as Blue\_Structure in Fig.~\ref{Figure1}). This bright structure could be a jet or a loop brightening, and is not detected in the red wing of H$\alpha$ or on any passband of the co-aligned AIA dataset. Distinct EB signatures are then observed at the loop foot-points where it connects to the solar surface (named EB1 and EB2 in Fig.~\ref{Figure1}, while the areas above the EBs are named as SEB1 and SEB2 respectively). A jet in the H$\alpha$ red wing is also seen (Red\_Jet in Fig.~\ref{Figure1}). The entire event occurs hidden below the chromospheric canopy, as seen in the H$\alpha$ line core. This supports the suggestion that these events are entirely upper photospheric/lower chromospheric. There is some evidence of a minor H$\alpha$ wing brightening at the approximate location of EB2 prior to the appearance of the incoming brightening (Blue\_Structure). However, this is only a 20\% increase compared to a local background profile and could not be classified as an EB. The EBs are observed for approximately 6 minutes but the observations end before their complete disappearance. The temporal evolution of the features was studied by integrating their intensity over the corresponding slice areas (right frames of Fig.~\ref{Figure1}). The integrated intensity from the slices is plotted in Fig.~\ref{Figure2} with intensity offsets applied to reduce overlap.

\begin{figure*}[!h]
\begin{center}
\plotone{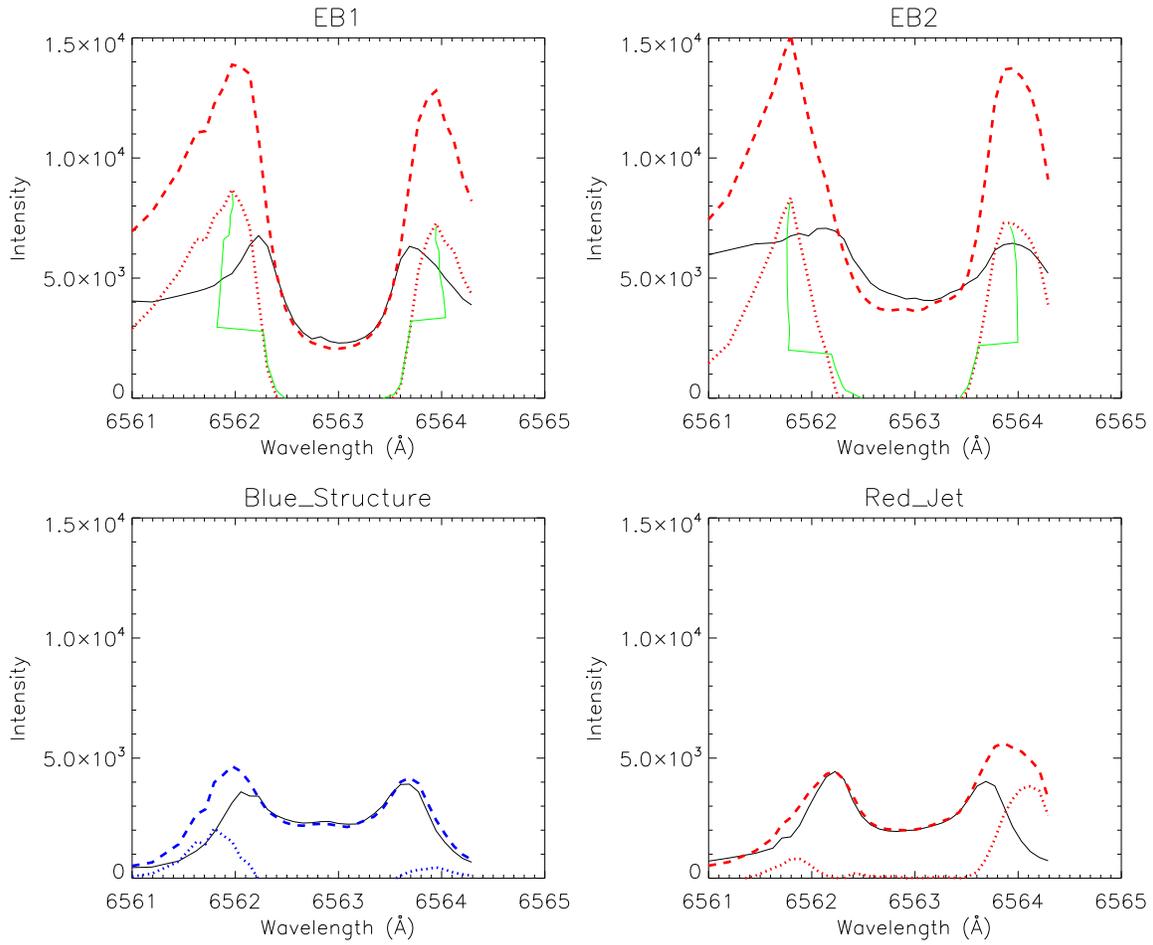}
\end{center}
\centering
\caption{H$\alpha$ profiles of the main features discussed in the text. The reference profile is the solid black line in each panel, with the dashed line showing the profile of the feature. The dotted line shows the residual profile of the feature with the reference profile removed. The green solid lines denote the bisectors for the residual red and blue wing enhancements associated with the EBs.}

\label{FigurePro}
\end{figure*}

\begin{figure*}[!h]
\begin{center}
\plotone{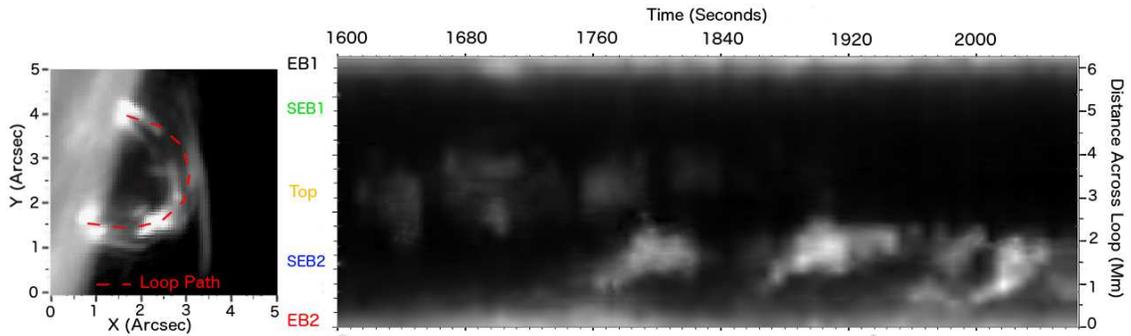}
\end{center}
\centering
\caption{The loop structure and time slice along the loop. The right panel shows the intensity of each point along the loop starting from the second, lower EB (EB\textbf{2}) and moving anti-clockwise. The vertical axis of the right hand panel is position along the loop structure. This slice was created at a blue position shift of -1.548~\AA\ relative to H$\alpha$ core. The left hand image shows the relative slice, taken at a frame of T=1800s.}

\label{Figure3}
\end{figure*}

The brightening in the blue wings of H$\alpha$ occurs first (Blue\_Structure in  Fig.~\ref{Figure1}, at time T=1560s from Movie1 and Fig.~\ref{Figure2}), and connects to the top of the loop, with an inflow velocity of about 50~km s$^{-1}$. This creates a brightening in  Loop\_Top, as evidenced in Fig.~\ref{Figure2} at T=1600s. This is then followed by the onset of the EBs at the loop foot-points approximately 1 minute later. The EB's brightness then expands to roughly 500~km above the foot-points (SEB1 and SEB2) after another 1 minute. The final event within this sequence is the appearance of a jet in the H$\alpha$ red wing at T=1752s. This red wing jet moves outward, with the top of the loop acting as its apparent source, with a velocity of approximately  of 65~km s$^{-1}$. The measured velocity is the magnitude of the small Doppler component seen in Fig.~\ref{FigurePro} of roughly 14~km s$^{-1}$ and the apparent transverse component.

The line profiles of EB1, EB2, Blue\_Structure and Red\_Jet are shown in Fig.~\ref{FigurePro}. The profiles were made by averaging over 9 pixels. The EB profiles were taken at  T=1712s, when the EBs first strongly appear. The times for the blue brightening and red jet are at T=1656s and T=1800s respectively corresponding to the peak in their intensities. The background profile varies between the different structures/positions due to the large variation of the canopy at the limb near the sunspot. The line profiles of the EBs were taken with reference to a background line profile of similar observation angle $\mu$. With these residual EB profiles, a bisector analysis was done of the blue and red intensity spikes related to the EBs  to find any non-thermal broadenings within the system, or general shifting of the line profile on both sides. Example results are shown in Fig.~\ref{FigurePro}. The results from the bisector analysis show negligible or no shift throughout the time series.

Fig.~\ref{Figure2} shows multiple rises in SEB2. To investigate this, a curve was fitted along the loop to study the time evolution within the structure (Fig.~\ref{Figure3}). The corresponding time-distance plot (right panel in Fig.~\ref{Figure3}) shows bursts of intensity which pulse along roughly the same point in the loop as the intensity slice SEB2 in Fig.~\ref{Figure1}. The burst velocity is of the order of 10 km s$^{-1}$, with repeated bursting every 2 minutes, with each burst lasting for over a minute. The intensity bursts are only seen in the blue wings of H$\alpha$, though in the red wings, the chromospheric jet is seen. The jet in the red wing of H$\alpha$ has a foot-point which is co-spatial and co-temporal with the top of the ejected material from EB2, with the bursting also first appearing at T=1752s.

EB2 has the morphology of a typical EB until the first bursting of material, which causes it  to have a `top-heavy' shape. The following bursts alter this shape, causing EB2 to have a more atypical upright Y morphology. While this is unusual to observe, the event begins as a regular EB in shape, size and brightness as compared with other studies \citep{Nelson, viss}, and evolves after eruptive bursting changes the morphology. EB1 shows no apparent bursting  and a typical morphology. 

\begin{figure*}[!t]
\begin{center}
\plotone{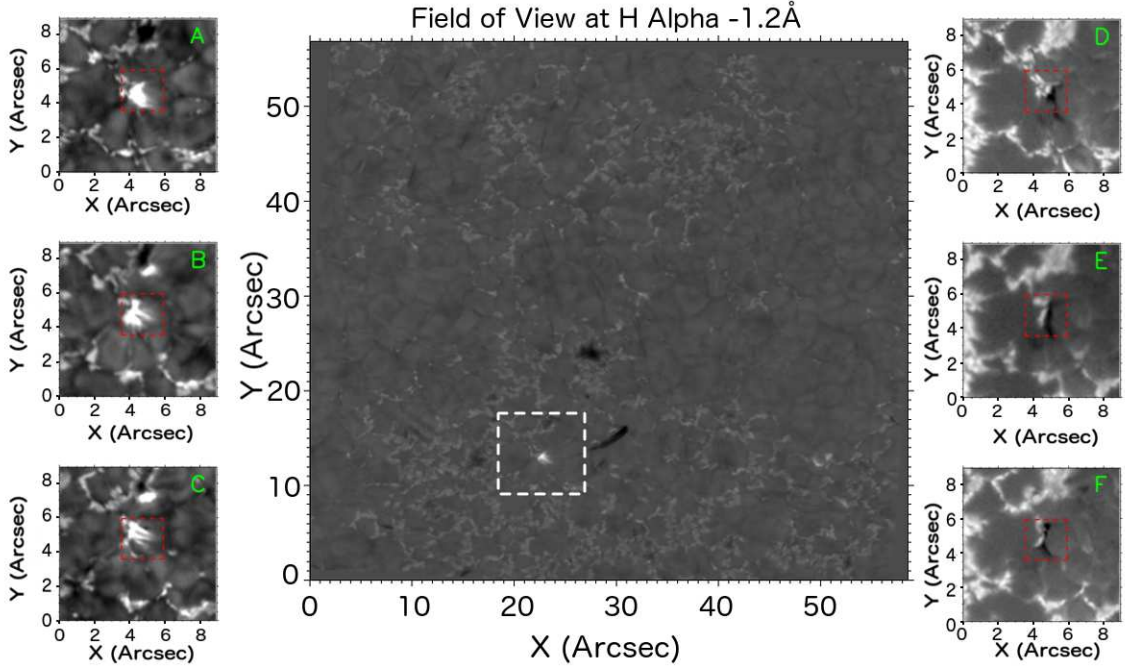}
\end{center}
\centering
\caption{Center: The FOV of the disk dataset with a white box showing the region of interest. Left column: H$\alpha$ -1.2~\AA\ images of the EB. Right column: Co-spatial and co-temporal Stokes-$V$ images. A, D: T = 480s; B, E: T = 760s; C, F: T = 1000s. Movie2 available online shows the temporal evolution of H$\alpha$ -1.2~\AA\ with Stokes-$V$ in greater detail.}
\label{Figure4}
\end{figure*}

Line-of-sight components of velocity could create a Doppler shift in the line profile at the sides of the loop connecting the two EBs in the limb dataset. If the loop was bent (i.e. part of the loop was in a plane perpendicular to the line of sight), the bursting seen in Fig.~\ref{Figure3} may have occurred on both sides of the loop, while only being observed in one. Alternatively, the material being ejected from the lower EB2 may have been dropped back down the other side of the loop. This is not seen in the red wing of H$\alpha$, or from the bisector analysis. However if material was falling with a Doppler component greater than 59 km s$^{-1}$, the red-shifted material will move outside the spectral domain of the H$\alpha$ scan and would not be detected.

\subsection{Ellerman Bomb on disk}

\begin{figure}[!h]
\begin{center}
\plotone{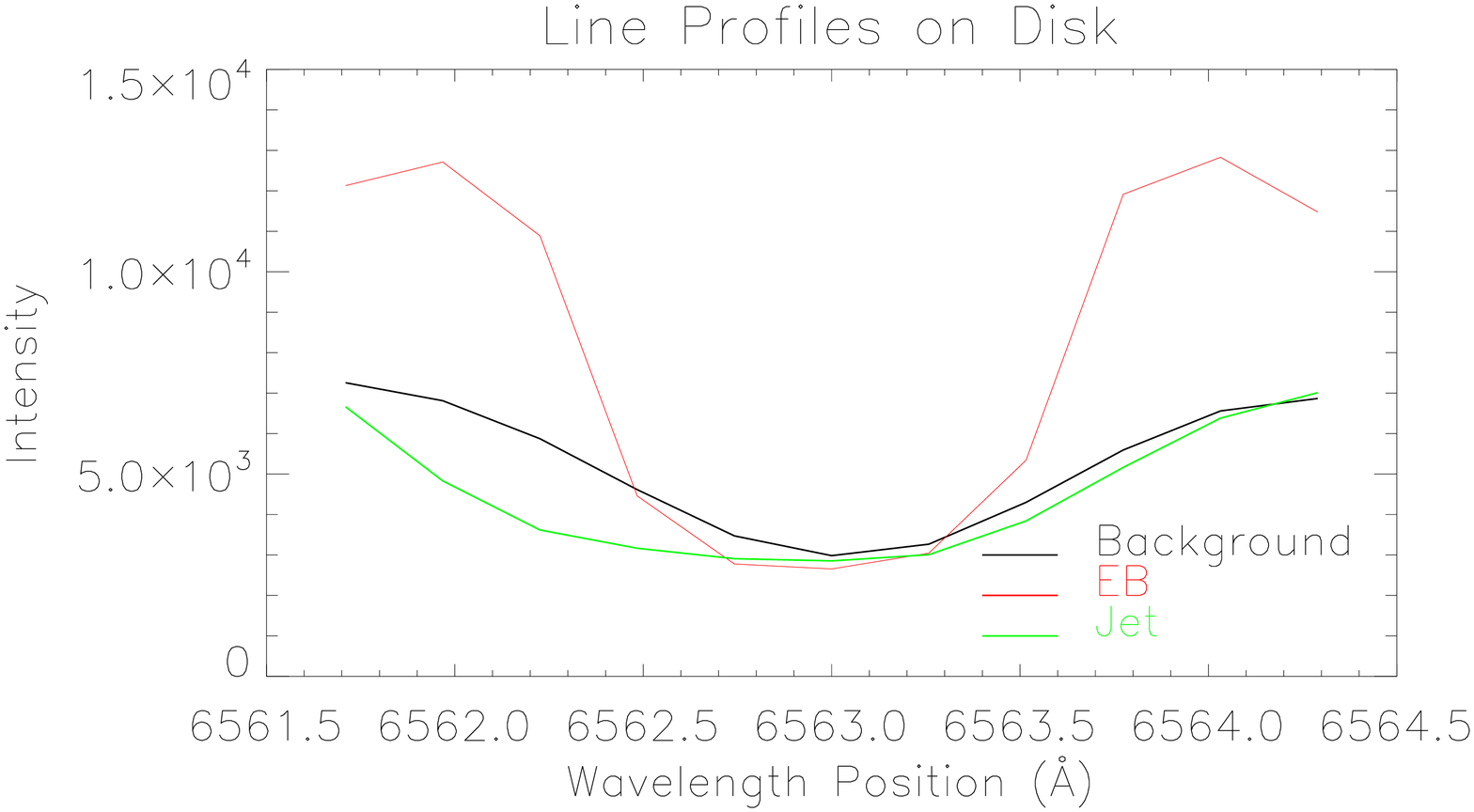}
\end{center}
\centering
\caption{H$\alpha$ profile of the Ellerman Bomb and an associated jet  from the disk dataset. A background profile is also shown. The profiles were taken at time T=352s.}
\label{FigureDisk}
\end{figure}

The disk observations show the occurrence of an Ellerman Bomb at the location of the inter-granular lanes. The lifetime of the EB is 13 minutes, while it's maximum size covers roughly 1 arcsec$^2$, with a peak brightness of 200\% of the local background in the wings of H$\alpha$. Snapshot images in H$\alpha$  core - 1.2~\AA\ and co-temporal Stokes-$V$ images are shown in Fig.~\ref{Figure4}. The H$\alpha$ line profile of the Ellerman Bomb is seen in Fig.~\ref{FigureDisk} along with the associated background profile. The profile of the Ellerman Bomb was gathered by averaging the H$\alpha$ intensity within a 40 pixel$^2$ region surrounding the EB (red boxes in Fig.~\ref{Figure4}), with the background profile being gathered from a 100 pixel$^2$ region of quiet sun from the same dataset. The Stokes-$V$ profiles show clear evidence for a bipolar small-scale magnetic structure, co-spatial with the classic EB signature of enhanced H$\alpha$ wings.  When the two opposite polarity fields first meet,  the H$\alpha$ wing brightness increases with a relatively round shape at the reconnection point, creating the EB. The Stokes-$V$ images  show that 5 minutes after the opposite polarities meet, the magnetic field has confined itself into the inter-granular lane. Interestingly, the simultaneous H$\alpha$ images show splitting of the EB brightening into two, with each flaring region moving along the inter-granular lane. The flaring region contains micro-jets, with a rising velocity of  $\sim$10 km s$^{-1}$, and a protrusion of 500~km. Two of the micro-jets are seen in panels B \& C in Fig.~\ref{Figure4} have a similar spatial rise-time as in the limb dataset.

\begin{figure}[!h]
\begin{center}
\plotone{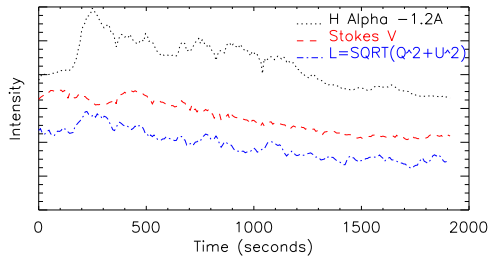}
\end{center}
\centering
\caption{Temporal evolution H$\alpha$ -1.2~\AA\ , Stokes-$V$, and L=$\sqrt{Q^{2}+U^{2}}$ over time at the EB on disk location. These lightcurves have been determined from a 40 pixel$^2$ area shown in panels A-F of Fig.~\ref{Figure4}, and a smaller 15 pixel$^2$ area for L, as to reduce noise as much as possible.}
\label{Figure6}
\end{figure}

In Fig.~\ref{Figure6} we show the linear polarisation (L=$\sqrt{Q^2 + U^2}$), circular polarisation (Stokes-$V$) and H$\alpha$ wing intensity at the EB location over time. The appearance of the EB corresponds to the peak in the H$\alpha$ wing intensity. This is co-temporal with an increase in L and a marginal decrease in Stokes-$V$, corresponding to a decrease in the line-of-sight magnetic field. This indicates that there is some tilting of the magnetic fields which is co-temporal with the appearance of the EB in H$\alpha$. The magnetic flux decreases over time in-line with a reduction in the H$\alpha$ wing intensity.

\begin{figure*}[!h]
\begin{center}
\plotone{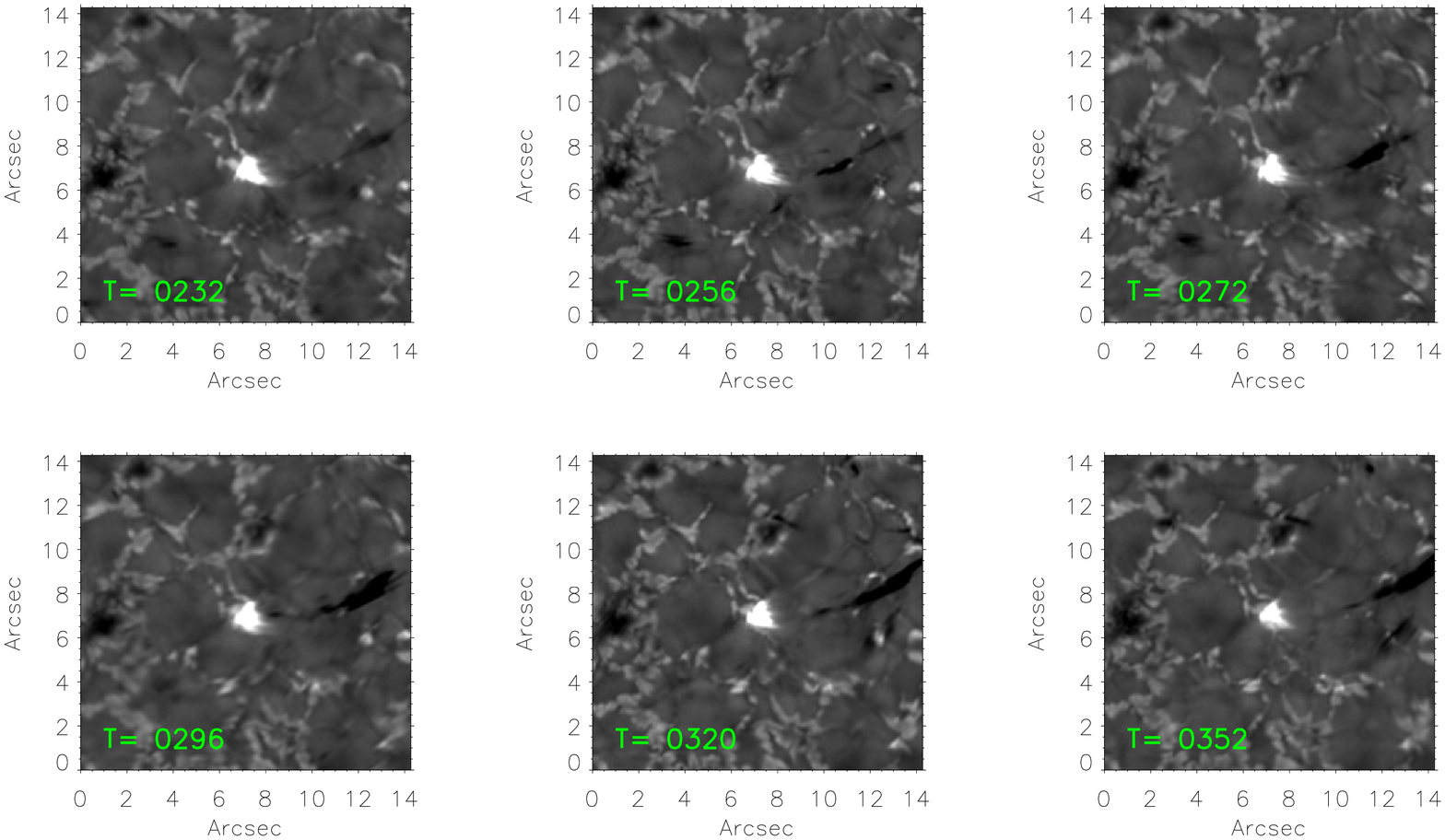}
\end{center}
\centering
\caption{H$\alpha$ -1.2~\AA\ imaging. A dark chromospheric jet is generated at the EB location. Time is in units of seconds.}
\label{Figure5}
\end{figure*}

Fig.~\ref{Figure5} shows clear evidence of a chromospheric jet emanating from the EB location, which is also shown in the center panel of Fig.~\ref{Figure4}. The line profile of the jet seen in Fig.~\ref{FigureDisk} created by averaging over a 3 pixel$^2$ sub-region within the jet where the jet appears strongest at T=352s. The weaker areas of the jet at this time also show the same features in the H$\alpha$ profile, only to a slightly lesser extent. The line-of-sight velocity of the jet was derived by measuring its Doppler velocity with respect to the reference H$\alpha$ line profile. As the jet follows a parabolic shape, its transverse velocity was determined by measuring the distance it travels with a curved space-time diagram. The space-time diagram indicates that the jet initiates with a large velocity and slows down over time. By combining the magnitude of the transverse velocity with the Doppler velocity, the absolute velocity of the jet could be estimated. The absolute velocity for the jet upon first appearing is 84 km s$^{-1}$, dropping to approximately 60 km s$^{-1}$ 90 seconds after its initiation. The drop in velocity is seen in both the  transverse and Doppler components, therefore the apparent retardation is not due to the curvature of the jet path. The drop in velocity could be due to gravity slowing down the jet over time.

\subsection{MURaM Simulations}

 We have also carried out numerical simulations using the MURaM radiative MHD code \citep{mu1}. The code has been extensively used for the study of small-scale photospheric phenomena \citep{Nelson1, shu, shel, chu}. MURaM solves the 3D radiative MHD equations on a Cartesian grid. A numerical grid of 480 x 480 pixels in the horizontal directions with 100 pixels in height, with corresponding physical dimensions of 12~Mm ($x$ axis) x 12~Mm ($z$ axis) x 1.4~Mm ($y$ axis), is used. The top boundary of the simulation box is closed for in-flows and out-flows, whereas the bottom boundary is open, and the side boundaries are periodic. The bottom 800~km is below the $\tau = 1$ line of 5000~\AA, while the top 600~km reaches into the photosphere. We note that this simulation does not have a chromosphere and cannot relate EBs to any chromospheric phenomena. The simulation is also quiet Sun, and so will not be able to reproduce the large EB intensities that are more likely to occur near active regions.

\begin{figure*}[!h]
\begin{center}
\plotone{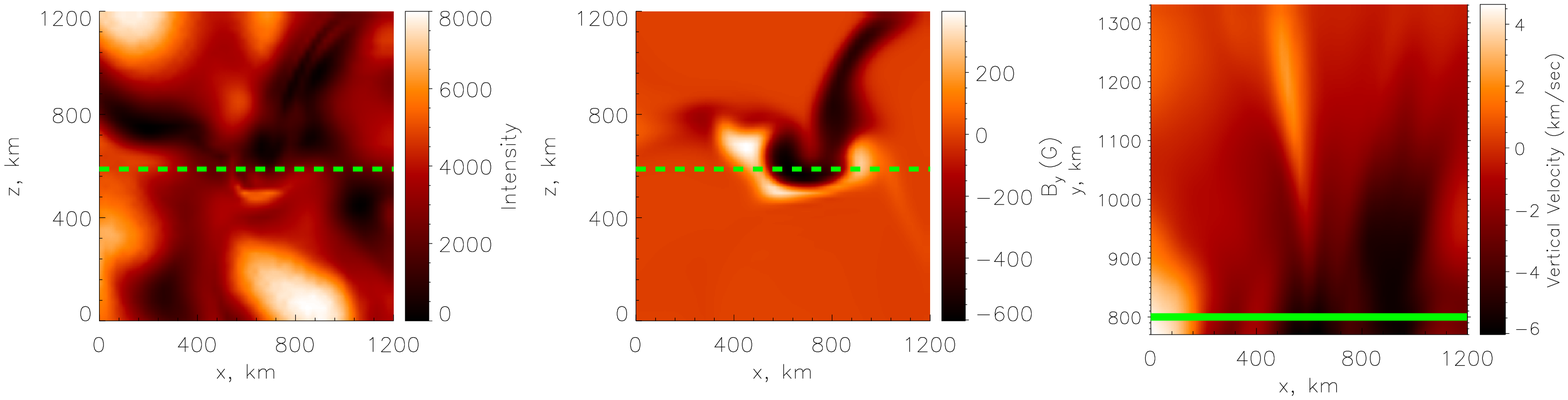}
\end{center}
\centering
\caption{Left: Computed H$\alpha$ intensity from the top of the simulated box at -1.2~\AA\ from line core. A horseshoe shaped brightening can be seen across the reconnection point between the 2 polarities in the center panel. Middle: The vertical magnetic field from the simulation. This cut was taken at a height of 300km above the $\tau = 1$ at 5000~\AA. Right: The vertical component of velocity from a cross section through the reconnection point. A jet can be seen propagating upward at the reconnection point. The dashed green lines denote the location of the vertical slice (right panel). The solid green line shows the approximate location of $\tau = 1$ at 5000~\AA.}
\label{Figure7}
\end{figure*}

The MURaM simulation used in this work has an initial bi-polar magnetic field with strength of $200~\mathrm{G}$ arranged in a checkerboard pattern with a $2~\mathrm{Mm}$ step. The simulation box was set to evolve for 50 minutes to allow the magnetic field to redistribute in the inter-granular lanes. After this initial stage, the snapshots containing the physical parameters of the plasma were written with a cadence of 2.5 seconds. The simulation is identical to \cite{Nelson1}, which detect a rise in temperature and H$\alpha$ wing intensity in a stronger reconnection site, closer to the brightness of an EB. Our findings are for a different event, where the rise in temperature along the reconnection line is accompanied by micro-jets, with velocities and vertical extent similar to our observations and previous work \citep{wat1}. The NICOLE line synthesis code \citep{socasnavarro} was used to generate H$\alpha$ line profiles in 1D non-LTE regime. \cite{leen} used a more intensive 3D approach to provide a  realistic model for the H$\alpha$ line core. We believe that for photospheric phenomena in the wings of H$\alpha$, such as EBs, the 1D approach is sufficient. \\

The output of the simulation shows a region in the inter-granular lane with a bi-polar magnetic field structure, shown in the centre panel of Fig.~\ref{Figure7}. The vertical velocity field along the reconnection point shows a small-scale jet which appears at the location where the opposite polarities meet, seen in the right panel of Fig.~\ref{Figure7}, with a velocity of  4-6 km s$^{-1}$, lasting $\sim$1 minute. A temperature rise is seen in this area, with the temperature rise being directly related to the increased emission in the wings of the simulated H$\alpha$ line (-1.2~\AA\ from line core). This curved brightened structure follows the rise in temperature seen across the reconnecting area, along with the shape of the jet structure in the vertical velocity field.\\

The rise in local temperature enhances the H$\alpha$ wings, an EB signature. The rise in the wings of the simulated H$\alpha$ profile is 110\% of the background reference profile. This would not classify as an EB in previous research \citep{viss, rutt} but we believe that the same physical processes are occurring, albeit on a smaller scale. The lower activity may be due to the fact that this is a quiet sun simulation, and the vertical extent of the simulated box is relatively small. An indepth comparison will require a simulation with a stronger magnetic field in the vicinity of an active region and a box with a larger vertical extent that includes a chromosphere. The event shows that micro-jets are seen in both the simulations and observations of photospheric magnetic reconnection. The jet structure does not appear to rise with height in the simulated H$\alpha$ profile due to the fact that the observational angle $\mu=0$ and the rising of the jet is almost vertical, as the jet follows the reconnection point upward. The cross sections of the simulated box clearly show that the rising motion of this material has an extent of 400-500~km.

\section{CONCLUDING REMARKS}

We study the formation and evolution of EBs both near solar disk centre and at the solar limb. The on-disk dataset shows that the evolution of the EB closely follows the area between opposite polarity photospheric magnetic fields. This fits the current theory that EBs are formed as a result of magnetic reconnection in the photosphere. The dataset used in our analysis also shows that the EB flares at a point when the transverse magnetic field increases, and the line of sight magnetic field decreases, suggesting that the magnetic fields tilt, forcing reconnection. The observations begin shortly before the EB is observed so it is difficult to determine the source of the bi-pole, which is reminiscent of bi-poles observed by \cite{viss, Nelson1}. A scenario where the tilting of the magnetic field forces a reconnection event within a bi-polar structure may fit with some topologies of photospheric reconnection, such as those caused by an emerging flux region within existing magnetic fields \citep{wat, hashi}, or reconnection of a resistive, undulatory ``sea serpent" flux emergence \citep{geo, par}.

The limb dataset shows a brightening in the blue wings of H$\alpha$ fuelling material and energy into a loop structure. Shortly after this occurs EBs flare at the foot-points of this loop structure. The idea presented here is that this fuelling of material into the loop structure and thus the loop foot-points, causes the flaring and detection of the EBs. This is analogous with reconnection in the upper atmosphere which has been observed to be due to in-flows on a much larger scale \citep{su}. This would be a new mechanism for the fuelling of EBs. The statement does not imply that incoming jet structures above the EB sites cause the reconnection but may instead fuel an existing site. In our limb dataset a small brightened structure existed at the loop foot-point prior to the appearance of blue wing brightening. Our bisector analysis did not reveal any activity on the other end of the loop.  This indicates that the bursting is fuelling the red jet. Although the bisector analysis shows no Doppler shift, this result should be treated with caution as any overlying fibril structure from the canopy could create asymmetries in the line profiles and cause erronous results. 

The limb dataset also shows that the entire event, including the red jet and the blue wing brightening connecting to the loop structure, occur below $\tau = 1$ at H$\alpha$ line core. The blue wing brightening appears to traverse horizontally across the solar surface, and so a Doppler scan in H$\alpha$ would not reveal this structure if viewed from disk centre. These structures would not be observable in H$\alpha$ line core from above either, as they would be beneath the canopy. This suggests that, although difficult to observe, jet associations with EBs may be a more common occurrence than previously reported.

In the disk dataset a jet was initially observed with a large velocity similar to \cite{roy}. The corresponding Doppler components are also similar to \cite{yang, wat1}. However, the retardation of the jet in both transverse and line of sight velocity in our study has not been observed before. The jet is seen approximately 80 seconds after the appearance of the EB. This is much less than the previous studies, with no distinct follow-up of a returning jet noted in previous works. The EB flares reaching peak intensity in a very short time (40 seconds, see Fig.~\ref{Figure4}). The threshold energy for the initiation of the jet may therefore occur earlier than previous reports. The lack of a red-shifted jet may be due to the material returning to the solar surface, following a trajectory that makes it fall on the other side of the loop, obscured from our field of view. In addition, it should be noted that the jet observed in the disk dataset appears to be more concentrated than in \cite{yang, wat1} as it propagates, while at the same time it is clearly distinguishable. 

Due to the similar formation velocities and geometry, we suggest that the rising of the EBs in the limb dataset with height is similar to the rising `micro-jet' events seen in the on-disk observations and simulations. This EB extension has been previously studied in detail by \cite{wat1} who note similar heights and velocities. However, the MURaM simulations show that this extension occurs between opposite polarities along the reconnection point. Although the micro-jet in the simulations is formed in an area of lower magnetic field strength, the velocity and size is roughly the same as that in the observations. We speculate that in an area of stronger reconnection, the energy may lead to increased localised heating with a more significant rise in intensity rather than into a rising velocity and size of the brightening. A detailed study that will account for varying the magnetic field strengths could address this issue. Magnetic flux sheets in the inter-granular lanes \citep{rear} can store magnetic energy between the granules, and may explain recently found extremely fast events observed in the wings of H$\alpha$ \citep{re1, re2}. These sheets may also be connected to the creation of EBs, as these are observed to be commonly located between granules. We suggest that the micro-jet acts as the flame of the candle, with the inter-granular lane brightening being the wick for the flame to burn down. Reconnection occurs, creating an EB, with the EB moving horizontally through the magnetic sheet in the inter-granular lane due to subsequent reconnection along the sheet.

In both observations, chromospheric jet connections to EBs have been observed, with the disk dataset showing a large area of inter-granular lane brightening around the EB ignition site, with a lot of stored magnetic energy for the jet to utilise. This conversion of magnetic energy stored within inter-granular lanes could then be enough to cause the creation of a chromospheric jet, driving energy upward from the photosphere. This also proposes an answer to why only some EBs form chromospheric jets, as it would depend on the local magnetic topology at the EB site, and how much energy is available. Another factor which may influence chromospheric jets forming from EBs has to do with the overlying magnetic canopy. This could be investigated further with a statistical sample of EBs that focuses on the minimum magnetic energy required for a chromospheric jet to emanate at the EB location.

\begin{acknowledgements}

We thank the anonymous referee for comments and suggestions that improved an earlier version of this paper. 
The Swedish 1-m Solar Telescope is operated on the island of La Palma by the Institute for Solar Physics of Stockholm University in the Spanish Observatorio del Roque de los Muchachos of the Instituto de Astrofísica de Canarias. We thank Peter S{\"u}tterlin, Johanna Vos, and Peter Halpin for assisting with the observations. We acknowledge support from Robert Ryans, Chris Smith, David Malone, and Gabriele Pierantoni with computing infrastructure. ES is a Government of Ireland Post-doctoral Research Fellow supported by the Irish Research Council. This research was supported by the SOLARNET project (www.solarnet-east.eu), funded by the European Commissions FP7 Capacities Program under the Grant Agreement 312495. Armagh Observatory is grant-aided by the N. Ireland Department of Culture, Arts and Leisure. AR would like to thank Armagh Observatory and Queen's University Belfast for funding. QUB and Armagh Observatory research is supported by the Science and Technology Facilities Council. This research was undertaken with the assistance of resources provided at the NCI National Facility systems at the Australian National University, supported by Astronomy Australia Limited, and at the Multi-modal Australian ScienceS Imaging and Visualisation Environment (MASSIVE) (www.massive.org.au). We thank the Centre 
for Astrophysics \& Supercomputing of Swinburne University of Technology (Australia) for the computational resources 
provided. S. Shelyag is the recipient of an Australian Research Council‚ Future Fellowship (project number FT120100057). The authors wish to acknowledge the DJEI/DES/SFI/HEA Irish Centre for High-End Computing (ICHEC) for the provision of computing facilities and support.

Facilities: SST (CRISP), SDO (AIA, HMI).

\end{acknowledgements}

\end{document}